\documentclass[hidelinks]{llncs}

\usepackage{times}
\usepackage{graphicx}
\usepackage{amssymb}
\usepackage{amsmath}
\usepackage{hyperref}
\usepackage{multirow}
\usepackage{graphicx}
\usepackage{subfig}
\usepackage{tikz}

\graphicspath{{figures/}}
\usepgflibrary{shapes.geometric}
\usetikzlibrary{shapes.geometric,shapes.misc,shapes}
\usetikzlibrary{arrows}
\usetikzlibrary{shadows}
\usetikzlibrary{fadings}
\usetikzlibrary{positioning}
\tikzset{vertex/.style={
	      circle,minimum size=6mm, very thick,draw=black!50, top
	      color=white,bottom color=black!20,inner sep=0pt}}

\let\phi\varphi
\let\epsilon\varepsilon

\renewcommand{\models}{\vDash}

\newcommand{\calB}{\mathcal{B}}

\newcommand{\calH}{\mathcal{H}}

\newcommand{\calT}{\mathcal{T}}

\newcommand{\algo}[1]{\ensuremath{\mathsf{#1}}}

\newcommand{\NP}{\ensuremath{\textsc{NP}}}

\newcommand{\SIGMA}[2]{\ensuremath{\Sigma_{\textrm{#1}}^{\textrm{#2}}}}

\newcommand{\tw}[1]{\mathit{tw}(#1)}

\newcommand{\dbdom}{\mathbf{C}}

\newcommand{\vardom}{\mathbf{V}}

\newcommand{\constants}[1]{\mathbf{#1}}
\newcommand{\variable}[1]{{#1}}
\newcommand{\variables}[1]{{\mathbf{#1}}}
\newcommand{\term}[1]{{#1}}
\newcommand{\terms}[1]{\mathbf{#1}}

\newcommand{\smods}[1]{\mathit{AS}(#1)}

\newcommand{\termt}{\term{t}}

\newcommand{\varW}{\variable{W}}
\newcommand{\varX}{\variable{X}}
\newcommand{\varY}{\variable{Y}}
\newcommand{\varZ}{\variable{Z}}
\newcommand{\varsW}{\variables{W}}
\newcommand{\varsX}{\variables{X}}
\newcommand{\varsY}{\variables{Y}}
\newcommand{\varsZ}{\variables{Z}}

\newcommand{\schema}[1]{\mathcal{#1}}

\newcommand{\relation}[1]{{\mathit{#1}}}

\newcommand{\atom}[1]{\underline{#1}}
\newcommand{\fullatom}[2]{{\relation{#1}(#2)}}
\newcommand{\domof}[1]{\mathit{dom}(#1)} 
\newcommand{\adomof}[1]{\mathit{adom}(#1)} 
\newcommand{\varof}[1]{\mathit{var}(#1)} 
\newcommand{\schof}[1]{\mathit{sch}(#1)} 

\newcommand{\schS}{\schema{S}}

\newcommand{\ground}[1]{\mathit{ground}(#1)}
\newcommand{\groundc}[2]{\mathit{ground}_{#1}(#2)}

\newcommand{\body}[1]{{\mathit{B}(#1)}}
\newcommand{\nbody}[1]{{\mathit{B}^-(#1)}}
\newcommand{\pbody}[1]{{\mathit{B}^+(#1)}}
\newcommand{\head}[1]{{\mathit{H}(#1)}}

\newcounter{cefalo}

\newcounter{cefalocont}

\newenvironment{changemargin}[2]{%
\list{}{\rightmargin#2\leftmargin#1
\parsep=0pt\topsep=0pt\partopsep=0pt}
\item[]}
{\endlist}

\newenvironment{indented}{\begin{changemargin}{1cm}{0cm}}{\end{changemargin}}

\def\qed{\hfill{\qedboxempty}      
  \ifdim\lastskip<\medskipamount \removelastskip\penalty55\medskip\fi}

\def\qedboxempty{\vbox{\hrule\hbox{\vrule\kern3pt
                 \vbox{\kern3pt\kern3pt}\kern3pt\vrule}\hrule}}

\def\qedfull{\hfill{\qedboxfull}   
  \ifdim\lastskip<\medskipamount \removelastskip\penalty55\medskip\fi}

\def\qedboxfull{\vrule height 4pt width 4pt depth 0pt}

\hypersetup{pdfinfo={
  Title={lpopt: A Rule Optimization Tool for Answer Set Programming}
  Author={Manuel Bichler, Michael Morak, and Stefan Woltran}
}}

\title{lpopt: A Rule Optimization Tool for Answer Set Programming}

\author{Manuel Bichler \and Michael Morak \and Stefan Woltran}

\institute{TU Wien, Vienna, Austria\\
  	   \{surname\}@dbai.tuwien.ac.at}

\begin{document}

\maketitle

\begin{abstract}
  State-of-the-art answer set programming (ASP) solvers rely on a program
  called a grounder to convert non-ground programs containing variables into
  variable-free, propositional programs. The size of this grounding depends
  heavily on the size of the non-ground rules, and thus, reducing the size of
  such rules is a promising approach to improve solving performance. To this
  end, in this paper we announce lpopt, a tool that decomposes large logic
  programming rules into smaller rules that are easier to handle for current
  solvers. The tool is specifically tailored to handle the standard syntax of
  the ASP language (ASP-Core) and makes it easier for users to write efficient
  and intuitive ASP programs, which would otherwise often require significant
  hand-tuning by expert ASP engineers. It is based on an idea proposed by Morak
  and Woltran (2012) that we extend significantly in order to handle the full
  ASP syntax, including complex constructs like aggregates, weak constraints,
  and arithmetic expressions. We present the algorithm, the theoretical
  foundations on how to treat these constructs, as well as an experimental
  evaluation showing the viability of our approach.
\end{abstract}

\section{Introduction}
\label{sec:introduction}

Answer set programming (ASP) \cite{iclp:GelfondL88,coll:MarekT99,%
cacm:BrewkaET11,book:GebserKKS12} is a well-established logic programming
paradigm based on the stable model semantics of logic programs. Its main
advantage is an intuitive, declarative language, and the fact that, generally,
each answer set of a given logic program describes a valid answer to the
original question. Moreover, ASP solvers---see e.g.\
\cite{ai:GebserKS12,lpnmr:AlvianoDFLR13,lpnmr:ElkabaniPS05,%
datalog:AlvianoFLPPT10}---have made huge strides in efficiency.

A logic program usually consists of a set of logical implications by which new
facts can be inferred from existing ones, and a set of facts that represent the
concrete input instance. Logic programming in general, and ASP in particular,
have also gained popularity because of their intuitive, declarative syntax. The
following example illustrates this:

\begin{example}\label{ex:intro}
  The following rule naturally expresses the fact that two people are relatives
  of the same generation up to second cousin if they share a great-grandparent.
  \vspace{-.5ex}
  \small
  \begin{verbatim}
uptosecondcousin(X, Y) :-
     parent(X, PX), parent(PX, GPX),
     parent(GPX, GGP), parent(GPY, GGP),
     parent(PY, GPY), parent(Y, PY), X != Y.
  \end{verbatim}
  \normalsize\vspace{-6ex}\qed
\end{example}

Rules written in an intuitive fashion, like the one in the above example, are
usually larger than strictly necessary. Unfortunately, the use of large rules
causes problems for current ASP solvers since the input program is grounded
first (i.e.\ all the variables in each rule are replaced by all possible, valid
combinations of constants). This grounding step generally requires exponential
time for rules of arbitrary size. In practice, the grounding time can thus
become prohibitively large. Also, the ASP solver is usually quicker in
evaluating the program if the grounding size remains small.

In order to increase solving performance, we could therefore split the rule in
Example~\ref{ex:intro} up into several smaller ones by hand, keeping track of
grandparents and great-grandparents in separate predicates, and then writing a
smaller version of the second cousin rule. While this is comparatively easy to
do for this example, this can become very tedious if the rules become even more
complex and larger, maybe also involving negation or arithmetic expressions.
However, since current ASP grounders and solvers become increasingly slower with
larger rules, and noting the fact that ASP programs often need expert
hand-tuning to perform well in practice, this represents a significant entry
barrier and contradicts the fact that logic programs should be fully
declarative: in a perfect world, the concrete formulation should not have an
impact on the runtime. In addition, to minimize solver runtime in general, it is
therefore one of our goals to enable logic programs to be written in an
intuitive, fully declarative way without having to think about various technical
encoding optimizations.

To this end, in this paper we propose the \verb!lpopt! tool that automatically
optimizes and rewrites large logic programming rules into multiple smaller ones
in order to improve solving performance. This tool, based on an idea proposed
for very simple ASP programs in \cite{iclp:MorakW12}, uses the concept of tree
decompositions of rules to split them into smaller chunks. Intuitively, via a
tree decomposition joins in the body of a rule are arranged into a tree-like
form. Joins that belong together are then split off into a separate rule, only
keeping the join result in a temporary atom. We then extend the algorithm to
handle the entire standardized ASP language \cite{web:aspcore}, and also
introduce new optimizations for complex language constructs such as weak
constraints, arithmetic expressions, and aggregates.

The main contributions of this paper are therefore as follows:
\vspace{-1ex}
\begin{itemize}
  \item we extend, on a theoretical basis, the \algo{lpopt}\ algorithm proposed
    in \cite{iclp:MorakW12} to the full syntax of the ASP language according to
    the ASP-Core-2 language specification~\cite{web:aspcore};

  \item we establish how to treat complex constructs like aggregates, and
    propose an adaptation of the decomposition approach so that it can split up
    large aggregate expressions into multiple smaller rules and expressions,
    further reducing the grounding size;

  \item we implement the \algo{lpopt}\ algorithm in C++, yielding the
    \verb!lpopt! tool for automated logic program optimization, and give an
    overview of how this tool is used in practice; and

  \item we perform an experimental evaluation of the tool on the encodings and
    instances used in the fifth Answer Set Programming Competition which show
    the benefit of our approach, even for encodings already heavily
    hand-optimized by ASP experts.
\end{itemize}

\section{Preliminaries}
\label{sec:preliminaries}

\paragraph*{General Definitions.} We define two pairwise disjoint countably
infinite sets of symbols: a set $\dbdom$ of \emph{constants} and a set $\vardom$
of \emph{variables}. Different constants represent different values
(\emph{unique name assumption}). By $\varsX$ we denote sequences (or, with
slight notational abuse, sets) of variables $\varX_1, \ldots, \varX_k$ with $k
\geqslant 0$. For brevity, let $[n] = \{1,\ldots,n\}$, for any integer $n
\geqslant 1$.

A (\emph{relational}) \emph{schema} $\schS$ is a (finite) set of
\emph{relational symbols} (or \emph{predicates}). We write $\relation{p}/n$ for
the fact that $\relation{p}$ is an $n$-ary predicate. A \emph{term} is a
constant or variable. An \emph{atomic formula} $\atom{a}$ over $\schS$ (called
\emph{$\schS$-atom}) has the form $\fullatom{p}{\terms{t}}$, where $\relation{p}
\in \schS$ and $\terms{t}$ is a sequence of terms. An \emph{$\schS$-literal} is
either an $\schS$-atom (i.e.\ a positive literal), or an $\schS$-atom preceded
by the negation symbol ``$\neg$'' (i.e.\ a negative literal). For a literal
$\ell$, we write $\domof{\ell}$ for the set of its terms, and $\varof{\ell}$ for
its variables. This notation naturally extends to sets of literals. For brevity,
we will treat conjunctions of literals as sets. For a domain $C \subseteq
\dbdom$, a (\emph{total} or \emph{two-valued}) \emph{$\schS$-interpretation} $I$
is a set of $\schS$-atoms containing only constants from $C$ such that, for
every $\schS$-atom $\fullatom{p}{\constants{a}} \in I$,
$\fullatom{p}{\constants{a}}$ is true, and otherwise false. When obvious from
the context, we will omit the schema-prefix.

A \emph{substitution} from a set of literals $L$ to a set of literals $L'$ is a
mapping $s: \dbdom \cup \vardom  \to \dbdom \cup \vardom$ that is defined on
$\domof{L}$, is the identity on $\dbdom$, and $\fullatom{p}{\term{t_1}, \ldots,
\term{t_n}} \in L$ (resp.\ $\neg \fullatom{p}{\term{t_1}, \ldots, \term{t_n}}
\in L$) implies $\fullatom{p}{s(\term{t_1}), \ldots, s(\term{t_n})} \in L'$
(resp., $\neg \fullatom{p}{s(\term{t_1}), \ldots, s(\term{t_n})} \in L'$).

\paragraph*{Answer Set Programming (ASP).} A \emph{logic programming rule} is a
universally quantified reverse first-order implication of the form
$$\calH(\varsX, \varsY) \gets \calB^+(\varsX, \varsY, \varsZ, \varsW) \wedge
\calB^-(\varsX, \varsZ),$$ where $\calH$ (the \emph{head}),  resp.\ $\calB^+$
(the \emph{positive body}), is a disjunction, resp. conjunction, of atoms, and
$\calB^-$ (the \emph{negative body}) is a conjunction of negative literals, each
over terms from $\dbdom \cup \vardom$. For a rule $\pi$, let $\head{\pi}$,
$\pbody{\pi}$, and $\nbody{\pi}$ denote the set of atoms occurring in the head,
the positive, and the negative body, respectively. Let $\body{\pi} = \pbody{\pi}
\cup \nbody{\pi}$. A rule $\pi$ where $\head{\pi} = \emptyset$ is called a
\emph{constraint}.  Substitutions naturally extend to rules. We focus on
\emph{safe} rules where every variable in the rule occurs in the positive body.
A rule is called $\emph{ground}$ if all its terms are constants. The grounding
of a rule $\pi$ w.r.t.\ a domain $C \subseteq \dbdom$ is the set of rules
$\groundc{C}{\pi} = \{ s(\pi) \mid s \text{ is a substitution, mapping }
\varof{\pi} \text{ to elements from } C \}$.

A \emph{logic program} $\Pi$ is a finite set of logic programming rules. The
schema of a program $\Pi$, denoted $\schof{\Pi}$, is the set of predicates
appearing in $\Pi$. The \emph{active domain} of $\Pi$, denoted $\adomof{\Pi}$,
with $\adomof{\Pi} \subset \dbdom$, is the set of constants appearing in $\Pi$.
A program $\Pi$ is ground if all its rules are ground. The \emph{grounding of a
program $\Pi$} is the ground program $\ground{\Pi} = \bigcup_{\pi \in \Pi}
\groundc{\adomof{\Pi}}{\pi}$. The \emph{(Gelfond-Lifschitz) reduct} of a ground
program $\Pi$ w.r.t.\ an interpretation $I$ is the ground program $\Pi^I = \{
\head{\pi} \gets \pbody{\pi} \mid \pi \in \Pi, \nbody{\pi} \cap I = \emptyset
\}$.

A $\schof{\Pi}$-interpretation $I$ is a \emph{(classical) model} of a ground
program $\Pi$, denoted $I \models \Pi$ if, for every ground rule $\pi \in \Pi$,
it holds that $I \cap \pbody{\pi} = \emptyset$ or $I \cap (\head{\pi} \cup
\nbody{\pi}) \neq \emptyset$, that is, $I$ satisfies $\pi$. $I$ is a
\emph{stable model} (or \emph{answer set}) of $\Pi$, denoted $I \models_s \Pi$
if, in addition, there is no $J \subset I$ such that $J \models \Pi^I$, that is,
$I$ is subset-minimal w.r.t.\ the reduct $\Pi^I$. The set of answer sets of
$\Pi$, denoted $\smods{\Pi}$, are defined as $\smods{\Pi} = \{ I \mid I \text{
is a } \schof{\Pi}\text{-interpretation, and } I \models_s \Pi \}$. For a
non-ground program $\Pi$, we define $\smods{\Pi} = \smods{\ground{\Pi}}$. When
referring to the fact that a logic program is intended to be interpreted under
the answer set semantics, we often refer to it as an \emph{ASP program}.

\paragraph*{Tree Decompositions.} A \emph{tree decomposition} of a graph $G =
(V,E)$ is a pair $\calT = (T, \chi)$, where $T$ is a rooted tree and $\chi$ is a
labelling function over nodes $t$ of $T$, with $\chi(t) \subseteq V$ called the
\emph{bag of $t$}, such that the following holds: (i) for each $v \in V$, there
exists a node $t$ in $T$, such that $v \in \chi(t)$; (ii) for each $\{v,w\} \in
E$, there exists a node $t$ in $T$, such that $\{v, w\} \subseteq \chi(t)$; and
(iii) for all nodes $r$, $s$, and $t$ in $T$, such that $s$ lies on the path
from $r$ to $t$, we have $\chi(r) \cap \chi(t) \subseteq \chi(s)$.  The
\emph{width} of a tree decomposition is defined as the cardinality of its
largest bag minus one. The \emph{treewidth} of a graph $G$, denoted by $\tw{G}$,
is the minimum width over all tree decompositions of $G$. To decide whether a
graph has treewidth at most $k$ is \NP-complete \cite{siamjadm:ArnborgCP87}. For
an arbitrary but fixed $k$ however, this problem can be solved (and a tree
decomposition constructed) in linear time \cite{siamcomp:Bodlaender96}.

Given a non-ground logic programming rule $\pi$, we let its \emph{Gaifman graph}
$G_\pi = (\varof{\pi}, E)$ such that there is an edge $(X, Y)$ in $E$ iff
variables $\varX$ and $\varY$ occur together in the head or in a body atom of
$\pi$. We refer to a tree decomposition of $G_\pi$ as a \emph{tree decomposition
of rule $\pi$}. The treewidth of rule $\pi$ is the treewidth of $G_\pi$.

\section{Rule Decomposition}
\label{sec:algorithms}

This section lays out the theoretical foundations for our rule decomposition
approach. First, we recall the algorithm from \cite{iclp:MorakW12}, and then
describe how it can be extended to handle three of the main extensions of the
ASP language, namely arithmetic expressions, aggregates, and weak constraints
(i.e.\ optimization statements), as defined in the ASP-Core language standard
\cite{web:aspcore}.

As demonstrated in Example~\ref{ex:intro}, rules that are intuitive to write
and read are not necessarily the most efficient ones to evaluate in practice.
ASP solvers generally struggle with rules that contain many variables since
they rely on a grounder-solver approach: first, the grounding of a logic
program is computed by a grounder. As per the definition in
Section~\ref{sec:preliminaries}, the size of the grounding can, in the worst
case, be exponential in the number of variables. For large rules, the grounding
step can already take a prohibitively large amount of time. However, the solver
is also adversely affected by this blowup. In practice, this leads to long
runtimes and sometimes the inability of the ASP system to solve a given
instance. This also contributes to the fact that, while the syntax of ASP is
fully declarative, writing efficient encodings still takes expert knowledge.

It is therefore desirable to have a way to automatically rewrite such large
rules into a more efficient representation. One way to do this is the rule
decomposition approach, first proposed in \cite{iclp:MorakW12}, which we will
briefly recall next.

\subsection{Decomposition of Simple Rules}

Generally speaking, the approach in \cite{iclp:MorakW12} computes the tree
decomposition of a rule, and then splits the rule up into multiple, smaller
rules according to this decomposition. While in the worst case this
decomposition may not change the rule at all, in practice it is often the case
that large rules can be split up very well. For instance, the large rule in
Example~\ref{ex:intro} will be amenable for such a decomposition.

Let us briefly recall the algorithm from \cite{iclp:MorakW12} which we will
refer to as the \algo{lpopt}\ algorithm. For a given rule $\pi$, the algorithm
works as follows:
\begin{enumerate}
  \item\label{decomp:step1} Compute a tree decomposition $\calT = (T, \chi)$ of
    $\pi$ with minimal width where all variables occurring in the head of $\pi$
    are contained in its root node bag. 

  \item\label{decomp:step2} For each node $n$, let $\mathit{temp}_n$ be a fresh
    predicate, and the same for each variable $\varX$ in $\pi$ and predicate
    $\mathit{dom}_X$. Let $\varsY_n = \chi(n) \cap \chi(p_n)$, where $p_n$ is
    the parent node of $n$. For the root node $\mathit{root}$, let
    $\mathit{temp}_\mathit{root}$ be the entire head of $\pi$, and, accordingly,
    $\varsY_\mathit{root} = \varof{\head{\pi}}$. Now, for a node $n$, generate
    the following rule:
    \begin{center}
      \begin{tabular}{l l l l l}
	$\mathit{temp}_n(\varsY_n)$ & $\leftarrow$ & \,\,\; $\{ \atom{a} \in
	\body{\pi}$ & $\mid$ & $\varof{\atom{a}} \subseteq \chi(n) \}$\\
	& & $\cup \, \{ \mathit{dom}_X(X)$ & $\mid$ & $\atom{a} \in \nbody{\pi}, X \in
	\varof{\atom{a}}, \varof{\atom{a}} \subseteq \chi(n),$\\
	& & & & $\not\exists \atom{b} \in \pbody{\pi}: \varof{\atom{b}}
	\subseteq \chi(n), \varX \in \varof{\atom{b}} \}$\\
	& & $\cup \, \{ \mathit{temp}_m(\varsY_m)$ & $\mid$ & $m \text{ is a child of }
	n\}.$
      \end{tabular}
    \end{center}

  \item\label{decomp:step3} For each $\varX \in \varof{\nbody{\pi}}$, for which
    a domain predicate $\relation{dom}$ is needed to guarantee safety of a rule
    generated above, pick an atom $\atom{a} \in \pbody{\pi}$, such that $\varX
    \in \varof{\atom{a}}$ and  generate a rule $$\fullatom{dom_\varX}{\varX}
    \leftarrow \atom{a}.$$
\end{enumerate}
Step~\ref{decomp:step3} is needed because splitting up a rule may make it
unsafe. In order to remedy this, a domain predicate is generated for each unsafe
variable that arises due to the rule splitting in step~\ref{decomp:step2}. The
following example illustrates how the algorithm works.

\begin{example}
  Given the rule $$\pi = \fullatom{h}{\varX, \varW} \gets \fullatom{e}{\varX,
  \varY}, \fullatom{e}{\varY, \varZ}, \neg \fullatom{e}{\varZ, \varW},
  \fullatom{e}{\varW, \varX},$$ a tree decomposition of $\pi$ could look as
  follows (note that we write in each bag of the tree decomposition not just the
  variables as per definition but also all literals of rule $\pi$ over these
  variables which is a more intuitive notation): 
  \begin{center}
    \begin{tikzpicture}
      \tikzstyle{every path}=[
	very thick,draw=green!50!black!50]
      \tikzstyle{every node}=[rectangle,
	    very thick,draw=green!50!black!50,
	    top color=white,bottom color=green!50!black!20]
      \tikzstyle{level 1}=[level distance=10mm, sibling distance=30mm]
      \tikzstyle{level 2}=[sibling distance=30mm]
      \tikzstyle{level 3}=[sibling distance=30mm]
      \node (root) {$\fullatom{h}{\varX, \varW}, \fullatom{e}{\varX, \varY},
		    \fullatom{e}{\varW, \varX}$}
      child {%
	node {$\fullatom{e}{\varY, \varZ} ,\neg \fullatom{e}{\varZ, \varW}$}
      };
    \end{tikzpicture}
  \end{center}
  Applying the \algo{lpopt}\ algorithm to $\pi$ with the tree decomposition
  above yields the following set of rules $\algo{lpopt}(\pi)$:
  $$\fullatom{dom_\varW}{\varW} \gets \fullatom{e}{\varW, \varX},$$
  $$\fullatom{temp}{\varY, \varW} \gets \fullatom{e}{\varY, \varZ}, \neg
  \fullatom{e}{\varZ, \varW}, \fullatom{dom_\varW}{\varW}, \text{ and}$$
  $$\fullatom{h}{\varX, \varW} \gets \fullatom{e}{\varX, \varY},
  \fullatom{e}{\varW, \varX}, \fullatom{temp}{\varY, \varW},$$ where
  $\relation{temp}$ is a fresh predicate not appearing anywhere else. \qed
\end{example}

Let $\Pi$ be a logic program. When the above algorithm is applied to all rules
in $\Pi$, resulting in a logic program $\algo{lpopt}(\Pi)$ as stated in
\cite{iclp:MorakW12}, the answer sets of $\Pi$ are preserved in the following
way: when all temporary atoms are removed, each answer set of
$\algo{lpopt}(\Pi)$ coincides with exactly one answer set from the original
program $\Pi$. Furthermore, the size of the grounding now no longer depends on
the rule size. In fact, it now only depends on the rule treewidth as the
following result states:

\begin{theorem}[\cite{iclp:MorakW12}]\label{thm:groundingsize}
  The size of $\ground{\algo{lpopt}(\Pi)}$ is bounded by $O(2^k \cdot n)$, where
  $n$ is the size of $\Pi$, and $k$ is the maximal treewidth of the rules in
  $\Pi$.
\end{theorem}

The above theorem implies that the size of the grounding of a program $\Pi$,
after optimization via the \algo{lpopt}\ algorithm, is no longer exponential in
the size of $\Pi$, but only in the treewidth of its rules. As
\cite{iclp:MorakW12} demonstrates, this decomposition approach already has a
significant impact on the size of the grounding in practical instances.

However, the ASP language standard \cite{web:aspcore} extends the ASP language
with other useful constructs that the \algo{lpopt}\ algorithm proposed in
\cite{iclp:MorakW12} cannot handle. These include arithmetic expressions,
aggregates, and weak constraints. Looking at concrete, practical instances of
ASP programs, e.g.\ the encodings used in recent ASP competitions
\cite{ai:CalimeriGMR16}, a large majority use such constructs. In the following,
we will therefore extend the \algo{lpopt}\ algorithm to be able to treat them in
a similar way.

\subsection{Treating Arithmetic Expressions}

Arithmetic expressions are atoms of the form $\varX = \varphi(\varsY)$, that is,
an equality with one variable (or constant number) $\varX$ on the left-hand
side, and an expression $\varphi$ on the right-hand side, where $\varphi$ is any
mathematical expression built using the variables from $\varsY$, constant
numbers, and the arithmetic connectives ``+,'' ``-,'' ``*,'' and ``/.'' In
addition to the positive and negative body, a rule $\pi$ may also contain a set
of such arithmetic expressions describing a relationship between variables with
the obvious meaning.

Clearly, in order to adapt the rule decomposition approach to this it is easy to
extend the definition of the graph representation of $\pi$ to simply contain a
clique between all variables occurring together in an arithmetic expression.
The \algo{lpopt}\ algorithm then works as described above up to
step~\ref{decomp:step2}. However, a problem may arise when, in
step~\ref{decomp:step3} of the \algo{lpopt}\ algorithm, a domain predicate
$\fullatom{dom_\varX}{\varX}$ is to be generated. Consider the following
example:

\begin{example}\label{ex:arithmetics}
  Let $\pi$ be the rule $\fullatom{a}{\varX} \gets \neg \fullatom{b}{\varX,
  \varY}, \fullatom{c}{\varY}, \fullatom{d}{\varZ}, \varX = \varZ + \varZ$. A
  simple decomposition according to the \algo{lpopt}\ algorithm may lead to the
  following rules: $$\fullatom{temp}{\varX} \gets \neg \fullatom{b}{\varX,
  \varY}, \fullatom{c}{\varY}, \fullatom{dom_\varX}{\varX}, \text{ and}$$
  $$\fullatom{a}{\varX} \gets \fullatom{d}{\varZ}, \varX = \varZ + \varZ,
  \fullatom{temp}{\varX}.$$ It remains to define the domain predicate
  $\relation{dom_\varX}$. According to the original definition of \algo{lpopt},
  we would get $$\fullatom{dom_\varX}{\varX} \gets \varX = \varZ + \varZ$$ which
  is unsafe. \qed
\end{example}

The conditions for safety of rules with arithmetic expressions are defined in
the ASP language specification \cite{web:aspcore}. As
Example~\ref{ex:arithmetics} shows, in order for such expressions to work with
the \algo{lpopt}\ algorithm a more general approach to defining the domain
predicates is needed in step~\ref{decomp:step3}. In fact, instead of choosing a
single atom from the rule body to generate the domain predicate, in general a
set of atoms and arithmetic expressions must be chosen. It is easy to see that
if a rule $\pi$ is safe then, for each variable $\varX \in \body{\pi}$, there
is a set $A$ of (positive) atoms and arithmetic expressions in the body of
$\pi$ that makes that variable safe. In step~\ref{decomp:step3} of the
\algo{lpopt}\ algorithm, for a variable $\varX$ we now choose such a set
$A_\varX$ of body elements in a greedy fashion as follows: let $\variables{S} =
\{ \varX \}$ the set of variables that we need to make safe. For each variable
$\variable{S} \in \variables{S}$, pick a (positive) atom from $\body{\pi}$ that
makes $\variable{S}$ safe, add it to $A_\varX$, and remove $\variable{S}$ from
$\variables{S}$. If no such atom exists in the body of $\pi$, greedily add the
smallest arithmetic expression $\variable{S} = \varphi(\varsY)$ in $\body{\pi}$
to $A_\varX$ and let $\variables{S} = \variables{S} \setminus \{ \variable{S} \}
\cup \varsY$. Repeat this process until $\variables{S}$ is empty. Since $\pi$
itself is safe and finite in size, the above procedure necessarily terminates.
Finally, generate the rule $\fullatom{dom_\varX}{\varX} \gets A_\varX$. It is
easy to see that this rule is safe and describes the possible domain of variable
$\varX$ as required. Note also that this rule can not be split up futher as
removing any single element of the rule would make it unsafe.

\begin{example}\label{ex:arithmetics:continued}
  A correct domain predicate for Example~\ref{ex:arithmetics} would be defined
  as follows: $$\fullatom{dom_\varX}{\varX} \gets \varX = \varZ + \varZ,
  \fullatom{d}{\varZ}.$$ This ensures the proper safety of all rules generated
  by the \algo{lpopt}\ algorithm. \qed
\end{example}

Note that the rule generated in Example~\ref{ex:arithmetics:continued} repeats
most of the atoms that the second rule generated in Example~\ref{ex:arithmetics}
already contains. It is not immediately obvious how such situations can be
remedied in general. Investigating this issue is part of ongoing work.

\subsection{Treating Weak Constraints}

As defined in \cite{web:aspcore}, a weak constraint $\pi [\term{k}:\terms{t}]$
is a constraint $\pi$ annotated with a term $\term{k}$ representing a weight
and a sequence of terms $\terms{t}$ occurring in $\pi$. The intended meaning is
that each answer set $I$ is annotated by a total weight $w(I)$, which is the sum
over all $\term{k}$ for each tuple of constants $\constants{c}$ that realize
$\terms{t}$ in $I$ and satisfy the body of $\pi$.  Such a weak constraint can
easily be decomposed by replacing $\pi [\term{k}:\terms{t}]$ with the rule $\pi'
= \fullatom{temp}{\term{k}, \terms{t}} \gets \body{\pi}$, where
$\relation{temp}$ is a fresh predicate, and the weak constraint $\bot \gets
\fullatom{temp}{\term{k}, \terms{t}} [\term{k}:\terms{t}]$. Finally, the
\algo{lpopt}\ algorithm is then applied to rule $\pi'$. This allows our rule
decomposition approach also to be applied in an optimization context (i.e.\
where the task for the solver is to find optimal answer sets w.r.t.\ their
weight).

\subsection{Treating Aggregate Expressions}

An aggregate expression, as defined in \cite{web:aspcore}, is an expression of
the form $$\termt \preccurlyeq \#\mathit{agg}\{ \terms{t} : \phi(\varsX) \},$$
where  $\termt$ is a term; $\preccurlyeq \, \in \{ <, \leqslant, =, \neq,
\geqslant, > \}$ is a builtin relation; $\mathit{agg}$ is one of $\mathit{sum}$,
$\mathit{count}$, $\mathit{max}$, and $\mathit{min}$; $\terms{t} = \langle
\termt_1, \ldots, \termt_n \rangle$ is a sequence of terms; and $\phi(\varsX)$
is a set of literals, arithmetic expressions, and aggregate expressions, called
the \emph{aggregate body}. Aggregates may appear in rule bodies, or recursively
inside other aggregates, with the following semantic meaning: Given an
interpretation $I$, for each valid substitution $s$ such that $s(\phi(\varsX))
\subseteq I$, take the tuple of constants $s(\terms{t})$. Let us denote this set
with $T$. Now, execute the aggregate function on $T$ as follows: for
$\#\mathit{count}$, calculate $|T|$; for $\#\mathit{sum}$, calculate
$\Sigma_{\terms{t} \in T} \termt_1$, where $\termt_1$ is the first term in
$\terms{t}$; for $\#\mathit{max}$ and $\#\mathit{min}$, take the maximum and
minimum term appearing in the first position of each tuple in $T$, respectively.
Finally, an aggregate expression is true if the relation $\preccurlyeq$ between
term $\termt$ and the result of the aggregate function is fulfilled.

Extending the \algo{lpopt} algorithm to aggregate expressions is again
straightforward: The rule graph $G_\pi = (V, E)$ of a rule $\pi$ containing
aggregate expressions is defined as follows: Let $V$ be the set of variables
occurring in $\pi$ outside of aggregate expressions. Let $E$ be as before and,
in addition, add, for each aggregate expression $\atom{e}$, a clique between all
variables $\varof{\atom{e}} \cap V$ to $E$. Intuitively, the rule graph should
contain, for each aggregate expression, a clique between all variables that
appear in the aggregate and somewhere else in the rule. Variables appearing
only in aggregates are in a sense ``local'' and are therefore not of interest
when decomposing the rule.

While the above transformation is straightforward, we can, however, go one step
further and also decompose the inside elements of an aggregate expression. To
this end, let $\termt \preccurlyeq \#\mathit{agg}\{ \terms{t} : \phi(\varsX,
\varsY) \}$ be an aggregate expression occurring in some rule $\pi$, where
$\varsX$ are variables that occur either in $\terms{t}$ or somewhere else in
$\pi$, and $\varsY$ are variables occurring inside the aggregate only. Replace
the aggregate expression with $\termt \preccurlyeq \#\mathit{agg}\{ \terms{t} :
\psi(\varsX, \varsZ), \fullatom{temp}{\terms{t}, \varsZ} \}$, and furthermore,
generate a rule $\fullatom{temp}{\terms{t}, \varsZ} \gets
\overline{\psi}(\varsY), \overline{\psi}_\mathit{dom}(\varsY)$, for some fresh
predicate $\relation{temp}$. Here, $\psi$ contains all those atoms from $\phi$
that contain a variable from $\varsX$, and $\overline{\psi}$ contains the rest.
$\overline{\psi}_\mathit{dom}$ contains domain predicates generated like in
step~\ref{decomp:step3} of the \algo{lpopt}\ algorithm, as needed to make the
temporary rule safe. The temporary rule can then be decomposed via \algo{lpopt}.
This is best illustrated by an example:

\begin{example}\label{ex:aggregates}
  Let $\pi$ be the following logic programming rule, saying that a vertex is
  ``good'' if it has at least two neighbours that, themselves, have a red
  neighbour: $$\fullatom{good}{\varX} \gets \fullatom{vertex}{\varX}, 2
  \leqslant \#\mathit{count}\{\varY : \fullatom{edge}{\varX, \varY},
  \fullatom{edge}{\varY, \varZ}, \fullatom{red}{\varZ} \}.$$ According to the
  above approach, the rule can now be split up as follows. Firstly, the
  aggregate is replaced: $$\fullatom{good}{\varX} \gets
  \fullatom{vertex}{\varX}, 2 \leqslant \#\mathit{count}\{\varY :
  \fullatom{edge}{\varX, \varY}, \fullatom{temp}{\varY} \},$$ and furthermore, a
  temporary rule is created as follows: $$\fullatom{temp}{\varY} \gets
  \fullatom{edge}{\varY, \varZ}, \fullatom{red}{\varZ}.$$ The latter rule is now
  amenable for decomposition via the \algo{lpopt} algorithm. \qed
\end{example}

Note that the above approach allows us to decompose, to a degree, even the
insides of an aggregate, which, for large aggregate bodies, can lead to a
further significant reduction in the grounding size.

\subsection{Correctness}

The correctness of the above extensions to the original algorithm follows by the
same arguments that prove the correctness of the original algorithm proposed in
\cite{iclp:MorakW12}, and trivially from the construction for arithmetic
expressions and safety. For the latter, note that for domain predicates of a
variable $\varX$ we explicitly select a set of atoms that make the variable
safe, and that such a set always exists, since the original rule is safe. For
the former two (namely weak constraints and aggregate expressions), the only
thing that needs to be examined is the first step: replacing (part of) the body
with a temporary predicate. But correctness of this is easy to see. Instead of
performing all joins within the weak constraint or aggregate, we perform the
join in a new, separate rule and project only relevant variables into a
temporary predicate. The weak constraint or aggregate then only needs to
consider this temporary predicate since, by construction, all other variables
not projected into the temporary predicate do not play a role w.r.t.\
optimization or aggregation. Finally, the original algorithm from
\cite{iclp:MorakW12} extended to handle arithmetic expressions, for which
correctness has already been established, is then applied to this new, separate
rule.

\subsection{Further Language Extensions}

The ASP-Core language specification \cite{web:aspcore}, as well as the
\emph{gringo} grounder\footnote{\url{http://potassco.sourceforge.net}}, allow
further constructs like variable pooling, aggregates with multiple bodies, or
with upper and lower bounds in the same expression, in addition to various
extensions that amount to syntactic sugar. These constructs make the above
explanations unnecessarily more tedious. However, from a theoretical point of
view, all of these additional constructs can be normalized to one of the forms
discussed in the previous subsections. Furthermore, as we shall see in the next
section, we have implemented the \algo{lpopt}\ algorithm to directly treat all
standard ASP language constructs and certain other additions, like variable
pooling. More details about this general approach, and the exact, but more
tedious, algorithm details, can be found in \cite{thesis:Bichler15}.

\section{Implementation}\label{sec:implementation}

A full implementation of the algorithm and its extensions described in
Section~\ref{sec:algorithms} is now available in the form of the \verb!lpopt!
tool, available with relevant documentation and examples at
\url{http://dbai.tuwien.ac.at/proj/lpopt}. The following gives a quick outline
of how to use the tool.

\verb!lpopt! accepts as its input any form of ASP program that follows the ASP
input language specification laid out in \cite{web:aspcore}. The output of the
program in its default configuration is a decomposed program that also follows
this specification. In addition, the tool guarantees that no language construct
is introduced in the output that was not previously present in the input (cf.\
Section~\ref{sec:algorithms}). Therefore, for example, a program without
aggregates will not contain any aggregates as a result of rule decomposition.
The following is a description of the parameters of the tool:

\begin{footnotesize}
\begin{verbatim}
Usage: lpopt [-idbt] [-s seed] [-f file] [-h alg] [-l file]
  -d        dumb: do not perform optimization
  -b        print verbose and benchmark information
  -t        perform only tree decomposition step
  -i        ignore head variables when decomposing
  -h alg    decomposition algorithm, one of {mcs, mf, miw (def)}
  -s seed   initialize random number generator with seed.
  -f file   the file to read from (default is stdin)
  -l file   output infos (treewidth) to file
\end{verbatim}
\end{footnotesize}
%

\noindent
In what follows, we will briefly describe the most important features of the
tool.

\paragraph{Tree Decomposition Heuristics.} As stated in
Section~\ref{sec:preliminaries}, computing an optimal tree decomposition w.r.t.\
width is an \NP-hard problem. We thus make use of several heuristic algorithms,
namely the \emph{maximum cardinality search (mcs)}, \emph{minimum fill (mf)},
and \emph{minimum induced width (miw)} approaches described in
\cite{iandc:BodlaenderK10}, that yield tree decompositions that provide good
upper bounds on the treewidth (i.e.\ on an optimal decomposition). It turns out
that in practice, since rules in ASP programs are usually not overly large,
these heuristics come close to, and often even yield, an optimal tree
decomposition for rules. The heuristic algorithm to use for decomposition can be
selected using the \verb!-h! command line parameter. Since these heuristic
approaches rely to some degree on randomization, a seed for the pseudo-random
number generator can be passed along with the \verb!-s! command line parameter.

\paragraph{Measuring the Treewidth of Rules.} Theorem~\ref{thm:groundingsize}
allows us to calculate an upper bound on the size of the grounding of the input
program. In order to do this, the maximal treewidth of any rule in an ASP
program must be known. The \verb!-l! switch of the \verb!lpopt! tool allows
this to be calculated. It forces the tool to perform tree decompositions on all
rules inside an input ASP program, simply outputting the maximal treewidth (or,
more accurately, an upper bound; see above) over all of them into the given
file, and then exiting.  Clearly, when a single ASP rule is given as input,
this switch will output a treewidth upper bound of that single rule.

\subsection*{Recommended Usage}

Assuming that a file \verb!enc.lp! contains the encoding of a problem as an ASP
program and that a file \verb!instance.db! contains a set of ground facts
representing a problem instance, the recommended usage of the tool is as
follows:
\begin{verbatim}
cat enc.lp instance.db | lpopt | grounder | solver
\end{verbatim}
In the above command, \verb!grounder! and \verb!solver! are programs for
grounding and for solving, respectively. One established solver that we will use
in the next section for our experimental evaluation is \emph{clasp}
\cite{ai:GebserKS12}. If \emph{clasp} is used as a solver together with the
\verb!lpopt! tool, we generally recommend the use of the \verb!--sat-prepro!
flag, which often speeds up the solving process substantially for decomposed
rules generated by \verb!lpopt! (by considering the fact that the truth values
of all temporary atoms generated by \verb!lpopt! are determined exactly by the
rule body, and need never be guessed).

\section{Experimental Evaluation}
\label{sec:evaluation}

We have tested our \verb!lpopt! tool and benchmarked the performance of
grounding and solving of programs preprocessed with \verb!lpopt! against
non-preprocessed ones. All benchmarks were made on the instance sets of the
fifth answer set programming competition 2014
\footnote{\url{https://www.mat.unical.it/aspcomp2014/}}, which, for most problem
classes, provides two encodings, one from 2013, and one from 2014.  The
benchmarks have been run on a 3.5GHz AMD Opteron Processor 6308 with 192 GB of
RAM to its disposal. We used the potassco software
suite\footnote{\url{http://potassco.sourceforge.net}}, namely \emph{gringo}
verison 4.5.3 as the grounder and \emph{clasp} version 3.1.3 as the solver. A
timeout of 300 seconds was set for solving, and 1000 seconds for grounding.
Furthermore, as suggested in the previous section, \emph{clasp} was called with
the \verb!--sat-prepro! flag enabled. In this paper, we will survey the most
important results. 

\paragraph*{Remark.} One central aim of our tool is to improve solving
performance for hand-written encodings by non-experts of ASP. In the spirit of a
truly declarative language, it shouldn't matter \emph{how} an encoding is
written as long as it is correct (i.e.\ w.r.t.\ runtime, there should not be a
difference between ``good'' and ``bad'' encodings). In this respect, the ASP
competition does not offer an optimal benchmark set since all encodings are
extensively hand-tuned by ASP experts. However, as to the best of our knowledge
there is no better-suited comprehensive benchmark set available, we will show
that even for these extensively hand-tuned ASP competition encodings our tool
can still find decompositions that decrease grounding size and improve solving
performance. However, there are also encodings that are so perfectly hand-tuned
that only trivial optimizations are possible with the current version of
\verb!lpopt!.

\paragraph*{Results.} Let us first note that the runtime of \verb!lpopt! itself,
for all encodings in the benchmark set, was always less than what can be
accurately measured on a computer system today. Applying our rule decomposition
algorithm thus comes virtually for free for hand-written encodings.  Out of the
49 encodings provided by the ASP competition, \verb!lpopt! was able to
syntactically rewrite 41 which indicates that, as mentioned above, even
extensively hand-tuned programs can be further decomposed in an automated
manner. The remaining eight encodings contained rules that were so small that no
further decomposition was possible (i.e.\ their Gaifman graph was a clique of
usually 3-4 nodes) and thus the output of \verb!lpopt! was the original,
unmodified encoding in these cases. In 27 of the 41 encodings rewritten by
\verb!lpopt!, the decompositions were trivial and had no significant impact on
the solving performance. This is due to the fact that only rules that were
already very small (and thus did not contribute much to the grounding size in
the first place) could be decomposed. In five cases out of the 41 rewritten
encodings, we noticed a decrease in solving performance (see the paragraph on
limitations of \verb!lpopt! below for an explanation) and in the remaining seven
cases, the \verb!lpopt! rewriting was able to speed up the solving process with
substantial improvements in three of these seven. Two of those were the stable
marriage problem encoding of 2013, and the permutation pattern matching encoding
of 2014 which we will take a closer look at below.  Full benchmark results for
the entire dataset can be found in \cite{thesis:Bichler15}.

\begin{figure}[t]
  \begin{multicols}{2}
	\includegraphics[width=\linewidth]{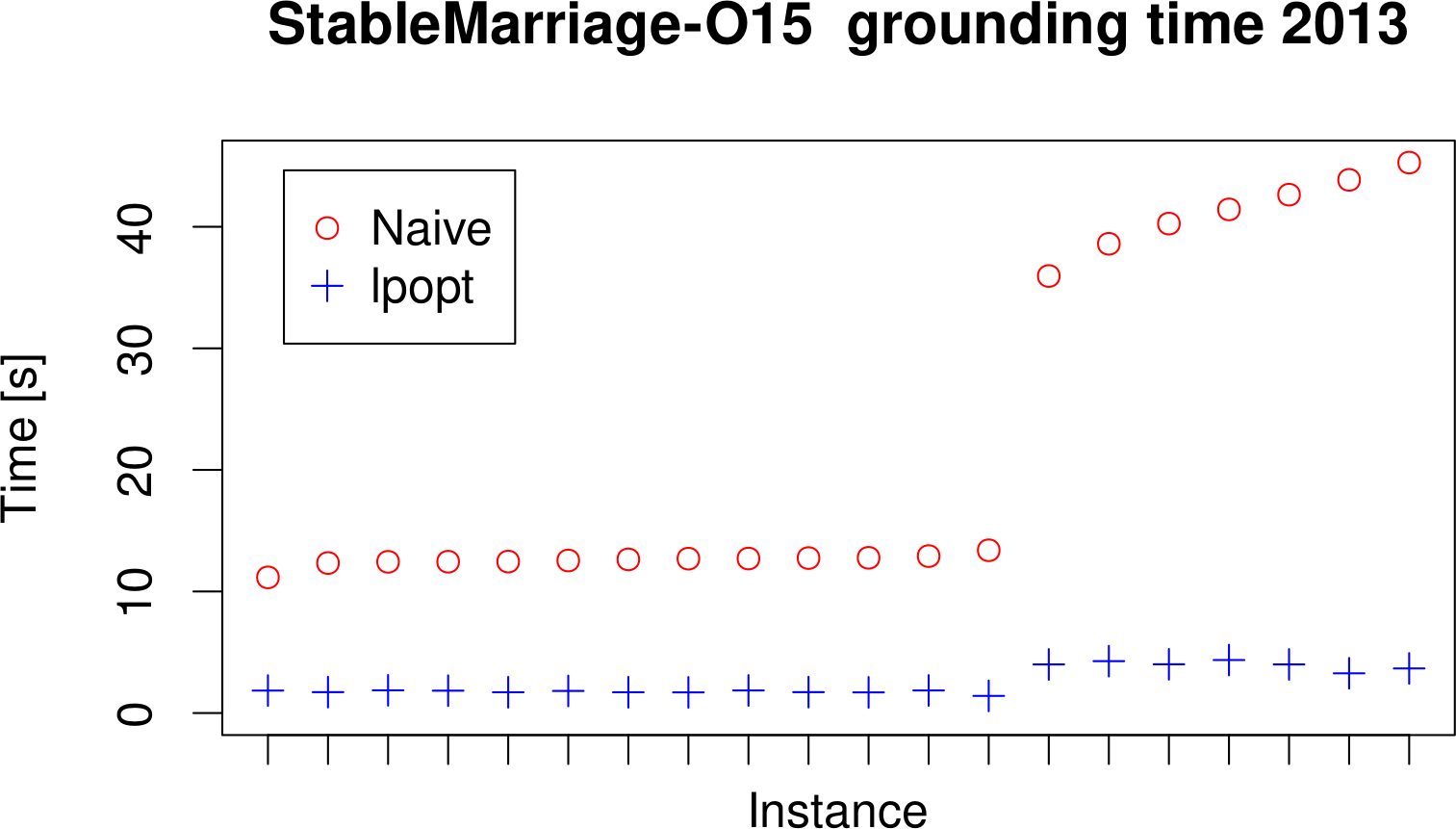}

	\vspace{-1.7ex}
	\begin{center} (a) \end{center}

	\includegraphics[width=\linewidth]{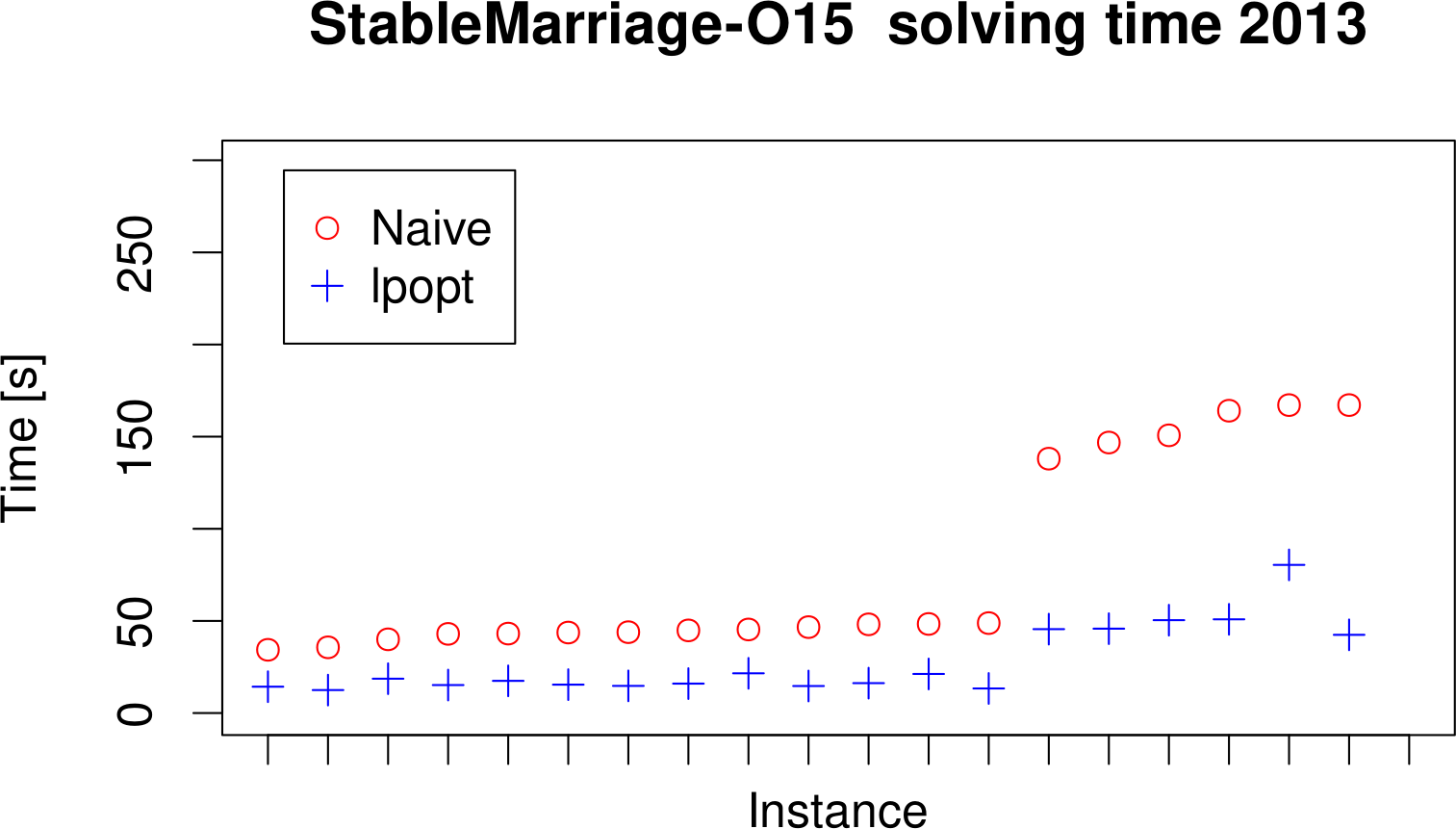}

	\vspace{-1.7ex}
	\begin{center} (b) \end{center}
  \end{multicols}
  \vspace{-3ex}
  \caption{Benchmark results for the stable marriage 2013 instances. The
  horizontal axis represents the individual test instances, sorted by runtime
  without rule decomposition.}
  \vspace{-3ex}
  \label{fig:stablemarriage}
\end{figure}

As can be seen in Figure~\ref{fig:stablemarriage}, both grounding and solving
time decrease dramatically. Notice that the grounding time is, in general,
directly correlated with the size of the respective grounding. With \verb!lpopt!
preprocessing, the grounding size decreases dramatically by a factor of up to
65. The grounder is thirty times faster when using preprocessing, and the solver
about three times. This is because of the following constraint in the encoding
that can be decomposed very well:

\small
\begin{verbatim}
:- match(M,W1), manAssignsScore(M,W,Smw), W1!=W,
     manAssignsScore(M,W1,Smw1), Smw>Smw1, match(M1,W),
     womanAssignsScore(W,M,Swm), womanAssignsScore(W,M1,Swm1),
     Swm>=Swm1.
\end{verbatim}
\normalsize

The constraint rule above is quite intuitive to read: There cannot be a man $M$
and a woman $W$, such that they would both be better off if they were matched
together, instead of being matched as they are (that is, to $W1$ and $M1$,
respectively). It encodes, precisely and straightforwardly, the condition of a
stable marriage. The 2014 encoding splits this rule up, making the encoding much
harder to understand. However, with \verb!lpopt! preprocessing, the grounding
and solving performance matches that of the hand-tuned 2014 encoding. This again
illustrates that the \algo{lpopt}\ algorithm allows for efficient processing of
rules written by non-experts that are not explicitly hand-tuned.

\begin{figure}
  \begin{multicols}{2}
	\includegraphics[width=\linewidth]{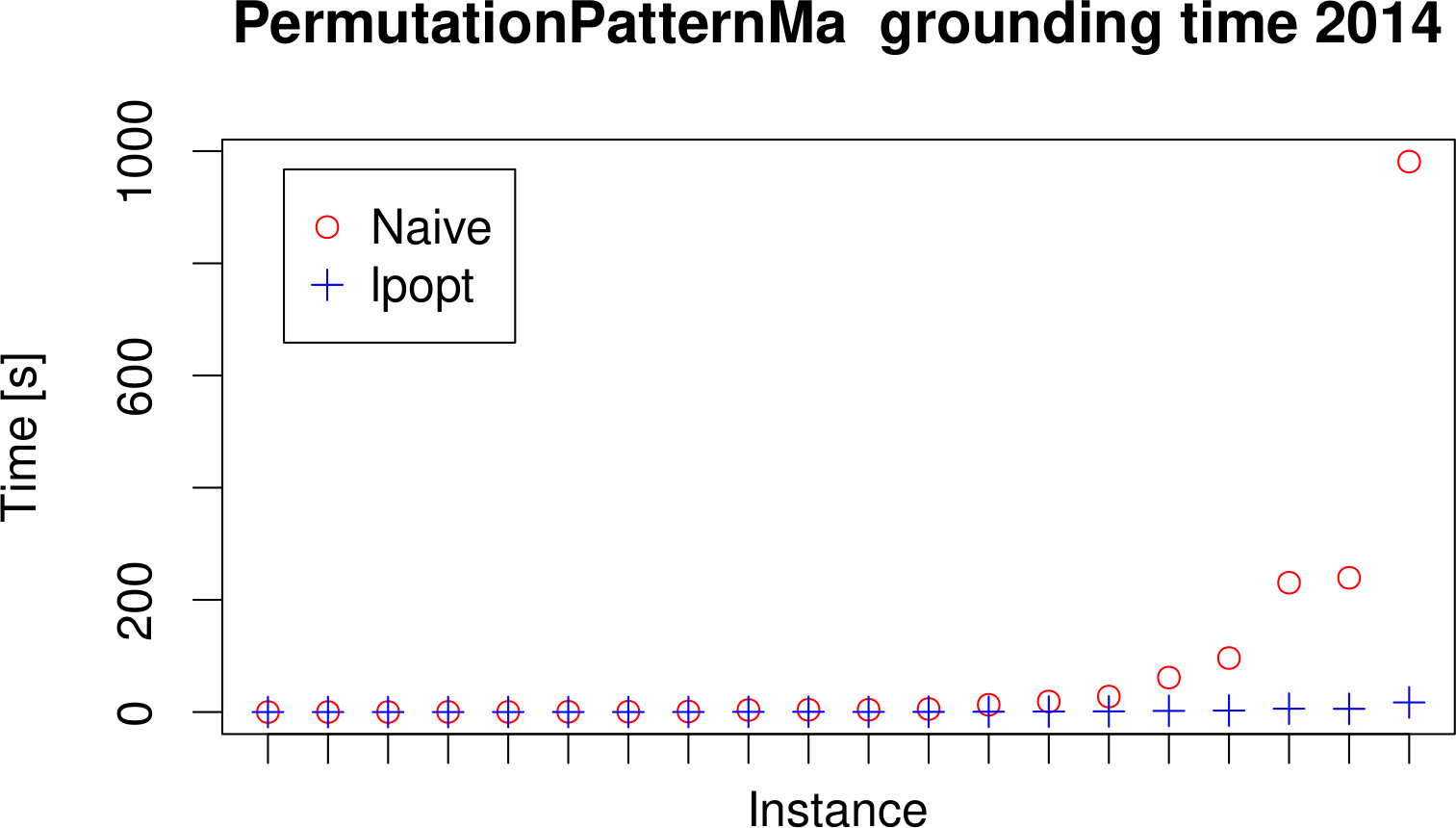}

	\vspace{-1.7ex}
	\begin{center} (a) \end{center}

	\includegraphics[width=\linewidth]{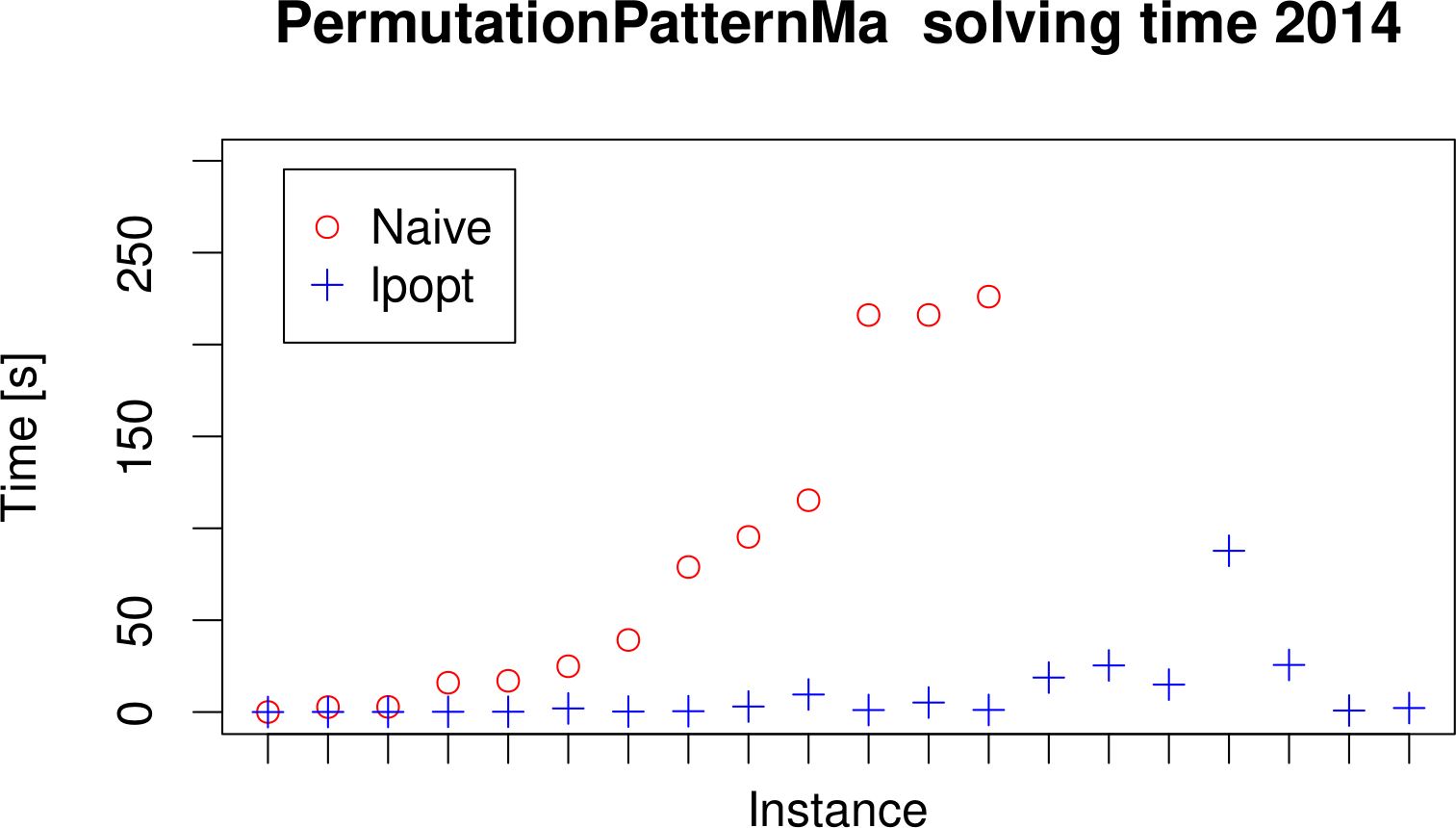}

	\vspace{-1.7ex}
	\begin{center} (b) \end{center}
  \end{multicols}
  \vspace{-3ex}
  \caption{Benchmark results for permutation pattern matching 2014. The
  horizontal axis represents the individual test instances, sorted by runtime
  without rule decomposition.}
  \vspace{-3ex}
  \label{fig:patternmatching}
\end{figure}

A second example of \verb!lpopt!'s capabilities is the permutation pattern
matching problem illustrated in Figure~\ref{fig:patternmatching}. The grounding
time of the largest instance is $980$ seconds without preprocessing and $17$
seconds with preprocessing. This instance was also impossible to solve within
the timeout window of $300$ seconds without \verb!lpopt! preprocessing, but
finishing within $88$ seconds when \verb!lpopt! was run first. 

\paragraph*{Other Use Cases.} \verb!lpopt! has also been employed in other works
that illustrate its performance benefits. In particular, several solvers for
other formalisms rely on a rewriting to ASP in order to solve the original
problem. Such rewritings can easily lead to the generation of large rules that
current ASP solving systems are generally unable to handle. For example, in
\cite{thesis:Heissenberger16} ASP rewritings for several problems from the
abstract argumentation domain, proposed in \cite{ecai:BrewkaW14}, are
implemented. In Section~4.6 of the thesis, the performance benefits of
\verb!lpopt! are clearly demonstated for these rewritings. Interestingly, these
rewritings also make heavy use of aggregates which goes to show that
\verb!lpopt! also handles these constructs well. Another example is
\cite{iclp:BichlerMW16}, where multiple rewritings for \SIGMA{P}{2}\ and
\SIGMA{P}{3}-hard problems are proposed and then benchmarked, again showcasing
that without \verb!lpopt! these rewritings could not be solved by current ASP
solvers in all but the most simple cases.

\paragraph*{Limitations.} However, we also want to point out some limitations of
the \algo{lpopt}\ algorithm. When a domain predicate is used by the algorithm,
the selection of atoms that generate this domain predicate is at the moment
essentially random, since the greedy selection depends on the order of the atoms
appearing in the rule. This approach, as discussed in
Section~\ref{sec:algorithms}, may thus not pick an optimal set of atoms.
However, it depends on this selection how many ground rules this domain
predicate rule will generate when passed to the grounder. Therefore, it may at
the moment be the case that the increased grounding size caused by the domain
predicate rules may destroy any benefit caused by splitting up the main rule.
This is precisely what caused the increase in solving time for the five
encodings out of 49 that \verb!lpopt! was able to rewrite but where solving
performance deteriorated.  Clearly, this begs the question of what the best
strategy is to select atoms to generate domain predicates. This is part of
ongoing work.

\section{Conclusions}
\label{sec:conclusions}

In this paper, we present an algorithm, based on a prototype from
\cite{iclp:MorakW12}, that allows the decomposition of large logic programming
rules into smaller ones that current state-of-the-art answer set programming
solvers are better equipped to handle. Our implementation handles the entire
ASP-Core-2 language \cite{web:aspcore}. Benchmark results show that in practice,
even for extensively hand-tuned ASP programs, our rule decomposition algorithm
can improve solving performance significantly. Future work will include
implementing this approach directly into state-of-the-art grounders like the
\emph{gringo} grounder used in our benchmarks, as well as further refining the
algorithm w.r.t.\ selection of domain predicate atoms, as discussed at the end
of Section~\ref{sec:evaluation}.

\paragraph*{Acknowledgments.} Funded by the Austrian Science Fund (FWF): Y698,
P25607.

\bibliographystyle{splncs03}
\bibliography{lopstr2016}

\end{document}